\documentclass[preprint,superscriptaddress,preprintnumbers,amsmath,amssymb,prd]{revtex4}
\usepackage{graphicx}

\begin{document}

\title{The effect of agglomeration of magnetic nanoparticles on the Casimir pressure
through a ferrofluid
}

\author{
G.~L.~Klimchitskaya}
\affiliation{Central Astronomical Observatory at Pulkovo of the
Russian Academy of Sciences, Saint Petersburg,
196140, Russia}
\affiliation{Institute of Physics, Nanotechnology and
Telecommunications, Peter the Great Saint Petersburg
Polytechnic University, Saint Petersburg, 195251, Russia}

\author{
V.~M.~Mostepanenko}
\affiliation{Central Astronomical Observatory at Pulkovo of the
Russian Academy of Sciences, Saint Petersburg,
196140, Russia}
\affiliation{Institute of Physics, Nanotechnology and
Telecommunications, Peter the Great Saint Petersburg
Polytechnic University, Saint Petersburg, 195251, Russia}
\affiliation{Kazan Federal University, Kazan, 420008, Russia}

\author{
E. N. Velichko}
\affiliation{Institute of Physics, Nanotechnology and
Telecommunications, Peter the Great Saint Petersburg
Polytechnic University, Saint Petersburg, 195251, Russia}

\begin{abstract}
The impact of agglomeration of magnetic nanoparticles on the Casimir pressure is
investigated in the configuration of two material plates and a layer of ferrofluid
confined between them. Both cases of similar and dissimilar plates are considered
in the framework of the Lifshitz theory of dispersion forces. It is shown that
for two dielectric (SiO${}_2$) plates, as well as for one dielectric (SiO${}_2$)
and another one metallic (Au) plates, an agglomeration of magnetite nanoparticles
results in only quantitative differences in the values of the Casimir pressure
if the optical data for Au are extrapolated to low frequencies by means of the
Drude model. If, however, an extrapolation by means of the plasma model is used
in computations, which is confirmed in experiments on measuring the Casimir
force, one finds that the pressure changes its sign when some share of magnetic
nanoparticles of sufficiently large diameter is merged into clusters by two or
three items. The revealed effect of sign change is investigated in detail at
different separations between the plates, diameters of magnetic nanoparticles
and shares of particles merged into clusters of different sizes. The obtained
results may be useful when developing ferrofluid-based microdevices and for
resolution of outstanding problems in the theory of Casimir forces.
\end{abstract}

\maketitle

\section{Introduction}

During the last few years much attention was given to ferrofluids which are
colloidal liquids consisting of magnetic nanoparticles suspended in some carrier
liquid (see, e.g., the monograph \cite{1} and Refs.~\cite{2,3,4,5,6,7,8,9}).
Ferrofluids are used in optical switches, optoelectronic communications,
mechanical and medical applications \cite{4,6,10,11,12}, and also in microdevices
playing a broad spectrum of roles \cite{13,14,15,16,17,18}. In the latter case,
ferrofluids may be confined in a narrow, submicrometer, gap between two
material plates. Under these conditions, the plates are subjected to the
Casimir force caused by the zero-point and thermal fluctuations of the
electromagnetic field \cite{19}. In the presence of a fluid in the gap, the
Casimir force may be both attractive and repulsive depending on materials
of the plates.

Nowadays the Casimir force is under active theoretical and experimental studies
(see Refs. \cite{20,21,22,23} for a review). Specifically, there is abundant
evidence that the Casimir force can be used in micro- and nanoelectromechanical
devices, such as Casimir oscillators, silicon chips, switches, optical
choppers etc. \cite{24,25,26,27,28,29,30,31,32}. This raises the question of
whether the Casimir force should be taken into account in ferrofluid-based
microdevices. In Ref. \cite{33} the Casimir pressure was investigated in the
case of magnetite nanoparticles suspended in kerosene or water between two
SiO${}_2$ plates. It was shown that an addition of a 5\% volume fraction of
magnetite nanoparticles leads to significantly different Casimir pressures as
compared to the case of nonmagnetic intervening liquid. It was found also that
at a fixed separation between the plates an addition to carrier liquid of
magnetite nanoparticles of some definite diameter does not influence the
Casimir pressure.

It has been known that magnetic particles suspended in a ferrofluid undergo
agglomeration which tends to diminish their surface energy \cite{1,6}.
To decrease an extent of agglomeration, magnetic nanoparticles are usually
coated with a surfactant which makes lower the surface tension between a
nanoparticle and a carrier liquid. However, even with a surfactant, some
share of magnetic nanoparticles merge into clusters composed of two, three
or more particles. The question arises: What is the effect of agglomeration
on the Casimir pressure between two material plates separated by a ferrofluid?

In this paper, we investigate the Casimir pressure through a ferrofluid under
different assumptions about nanoparticle diameter, materials of the plates,
and the extent of agglomeration of magnetite nanoparticles. It is shown
that the extent of agglomeration affects the pressure in a nontrivial way
depending on the other parameters of the problem and, what is more important,
on the theoretical approach used in calculations. Specifically, for two
similar dielectric plates separated by a ferrofluid the effect of agglomeration
of magnetite nanoparticles leads to only relatively small quantitative
differences in the Casimir pressure. The most interesting results are found
for a ferrofluid confined between two dissimilar plates, one metallic (Au) and
another one dielectric (SiO${}_2$). This is an example of three-layer systems
much studied in the case of nonmagnetic intervening layer \cite{19,22},
where the pressure can be repulsive. According to our results, for magnetite
nanoparticles of 10~nm diameter suspended in water as a carrier liquid
agglomeration leads to only relatively small variations in the values of
repulsive Casimir pressure. However, for nanoparticles of 20 nm diameter the
effect of agglomeration on the pressure appears essentially dependent on the
used model of the low-frequency dielectric response of Au.

It is the subject of considerable literature that the Lifshitz theory of
the van der Waals and Casimir forces agrees with the measurement results
only if the available optical data of a plate metal are extrapolated down
to zero frequency by means of the lossless plasma model. If the lossy
Drude model is used for extrapolation, the theoretical predictions are
excluded by the experimental data at the highest confidence level (see
Refs.~\cite{19,20,23,34,34a,34b} for a review and more recent experiments
\cite{35,36,37,38,39,40}). This result received the name Casimir puzzle
because it implies that the dielectric response of a metal to the
low-frequency fluctuating field is not that which is normally expected
to an ordinary electromagnetic field.

Specifically, we demonstrate that for magnetite nanoparticles of 20 nm
diameter the effect of agglomeration on the Casimir pressure remains again
entirely quantitative and relatively small if the optical data of Au are
extrapolated by means of the lossy Drude model. If, however, the
experimentally consistent lossless plasma model is used for extrapolation,
the agglomeration of magnetite nanoparticles results in the change of sign
of the Casimir pressure from repulsion to attraction at some separation
distance between the plates. The physical explanation to this effect
is provided. The transition conditions of the Casimir pressure from
repulsion to attraction under an impact of agglomeration are investigated
as functions of the share of nanoparticles merged into clusters and of
nanoparticle diameter. Possible applications of the obtained results are
discussed.

The paper is organized as follows. In Sec.~II, we present the Lifshitz
formula for the Casimir pressure adapted for a configuration of two
dissimilar plated separated by a ferrofluid. Section III contains the
computational results on the impact of agglomeration of magnetite
nanoparticles on the Casimir pressure for both dissimilar and similar
plates obtained using different theoretical approaches. In Sec.~IV, the
conditions for a change of sign of the Casimir pressure are investigated
for different shares of nanoparticles merged into clusters and different
nanoparticle diameters. Section V contains our conclusions and a
discussion.
%%%%%%%%%%%%%%%%%%%%%%%%%%%%%%%%%%%%%%%%%%%%%%%%%%%%%%%%%%%%%%%%%%%%%%%%%%%

\section{The Lifshitz formula for two dissimilar plates separated
by a ferrofluid}

We consider the configuration of two parallel nonmagnetic plates described by the
dielectric permittivities $\varepsilon^{(1)}(\omega)$ and $\varepsilon^{(2)}(\omega)$.
The gap between the plates of thickness $a$ is filled with a ferrofluid described by the
dielectric permittivity $\varepsilon_{\rm ff}(\omega)$ and magnetic permeability
$\mu_{\rm ff}(\omega)$. Material plates can be considered as semispaces if they are
thicker than 100~nm \cite{19} and $2~\mu$m \cite{41} in the case of metallic and
dielectric materials, respectively. The Casimir pressure at temperature $T$ can be
conveniently expressed in terms of the dimensionless variables by the following
Lifshitz formula:
\begin{eqnarray}
&&
  {P}(a)=-\frac{k_BT}{8{\pi}a^3}\sum_{l=0}^{\infty}
\vphantom{\sum}^{'}\int_{\sqrt{\varepsilon_{{\rm ff},l}\mu_{{\rm ff},l}}\zeta_l}^{\infty}
y^2dy
\label{eq1}\\
&&~~~
  \times \sum_{\alpha}\left[
\frac{e^y}{r_{\alpha}^{(1)}(i{\zeta}_l, y)r_{\alpha}^{(2)}(i{\zeta}_l, y)}-1\right]^{-1}.
  \nonumber
\end{eqnarray}
\noindent
Here, $k_B$ is the Boltzmann constant, the dimensionless Matsubara frequencies $\zeta_l$
are expressed via the dimensional ones $\xi_l$ by
\begin{equation}\label{eq2}
  \zeta_l=\frac{\xi_l}{\omega_{\rm cr}}=\frac{2a\xi_l}{c}
  =\frac{4\pi ak_BTl}{\hbar c}, \quad
  l=0,\,1,\,2,\,\ldots,
\end{equation}
\noindent
and all dielectric permittivities and magnetic permeability are calculated at these
frequencies along the imaginary frequency axis
\begin{eqnarray}
&&
\varepsilon_{{\rm ff},l}\equiv\varepsilon_{\rm ff}(i\xi_l)
=\varepsilon_{\rm ff}(i\omega_{c}\zeta_l),
\nonumber\\
&&
\varepsilon_{l}^{(k)}\equiv\varepsilon^{(k)}(i\xi_l)
=\varepsilon^{(k)}(i\omega_{c}\zeta_l),
\quad k=1,\,2,
\nonumber\\
&&
\mu_{{\rm ff},l}\equiv\mu_{\rm ff}(i\xi_l)
=\mu_{\rm ff}(i\omega_{c}\zeta_l).
\label{eq3}
\end{eqnarray}

The prime on the first summation sign in Eq.~(\ref{eq1}) means that the term with $l=0$
is divided by $2$, and the summation in $\alpha$ is over
two independent polarizations of the electromagnetic
field, transverse magnetic ($\alpha={\rm TM}$) and transverse electric
($\alpha={\rm TE}$).
Finally, the reflection coefficients on the first and second plates
 are given by
\begin{eqnarray}
&&
r_{\rm TM}^{(k)}(i\zeta_l,y)=\frac{\varepsilon_l^{(k)}y-\varepsilon_{{\rm ff},l}\sqrt{y^2+
(\varepsilon_l^{(k)}-\varepsilon_{{\rm ff},l}\mu_{{\rm ff},l})\zeta_l^2}}{\varepsilon_l^{(k)}y+
\varepsilon_{{\rm ff},l}\sqrt{y^2+(\varepsilon_l^{(k)}-\varepsilon_{{\rm ff},l}\mu_{{\rm ff},l})\zeta_l^2}},
\nonumber\\
&&
r_{\rm TE}^{(k)}(i\zeta_l,y)=\frac{y-\mu_{{\rm ff},l}\sqrt{y^2+(\varepsilon_l^{(k)}-
\varepsilon_{{\rm ff},l}\mu_{{\rm ff},l})\zeta_l^2}}{y+\mu_{{\rm ff},l}\sqrt{y^2+
(\varepsilon_l^{(k)}-\varepsilon_{{\rm ff},l}\mu_{{\rm ff},l})\zeta_l^2}}.
\label{eq4}
\end{eqnarray}

Now calculation of the Casimir pressure through a ferrofluid can be performed by
Eqs.~(\ref{eq1})--(\ref{eq4}) if one knows the values of dielectric permittivities
of the plates and ferrofluid at the pure imaginary frequencies,
$\varepsilon_l^{(k)}$, $\varepsilon_{{\rm ff},l}$, and of ferrofluid magnetic
permeability $\mu_{{\rm ff},l}$. Note, that the magnetic permeability quickly
decreases with increasing frequency and at room temperature becomes equal to unity
at $\xi\ll\xi_1$. For this reason, the magnetic properties of a ferrofluid,
as well as of any other magnetic body, influence the Casimir force only through the
term of Eq.~(\ref{eq1}) with $l=0$ \cite{42}.

Below we consider Au and SiO$_2$ plates with an intervening layer of water-based
ferrofluid containing $\Phi=0.05$ volume fraction of magnetite nanoparticles.
The dielectric permittivity of Au along the imaginary frequency axis was obtained
using the optical data of Ref.~\cite{43} extrapolated to lower frequencies by either
the plasma or the Drude model and repeatedly used in the literature
\cite{19,20,34,34a,34b,35,36,39,40}. At low frequencies the respective permittivities
behave as $\varepsilon_p^{(1)}\sim \omega_p^2/\xi^2$ and
$\varepsilon_D^{(1)}\sim \omega_p^2/(\gamma\xi)$, where $\omega_p$ is the plasma
frequency and $\gamma$ is the relaxation parameter of Au.

The permittivities of SiO$_2$ and water are taken from Refs.~\cite{44,45},
respectively. Specifically, for SiO$_2$ one has $\varepsilon^{(2)}(0)=3.801$.
An analytic expression for the dielectric permittivity of magnetite was
found in Ref.~\cite{33}  using the measured optical data of Ref.~\cite{46} and
the Kramers-Kronig relations. Combining the permittivities of water and magnetite
with the help of Rayleidh's mixing formula, the permittivity of ferrofluid
$\varepsilon_{{\rm ff},l}$ with a given volume fraction of nanoparticles $\Phi=0.05$
was obtained \cite{33}. With omitted conductivity of magnetite at low frequencies
(see Refs.~\cite{19,20,48,49,50} for the reasons why this option is more realistic
in computations of the Casimir force), one arrives at
$\varepsilon_{{\rm ff}}(0)=77.89$.

The resulting permittivities of Au (in two variants), SiO$_2$ and of a ferrofluid
are used in Secs,~III and IV to investigate an impact of agglomeration of
nanoparticles on the Casimir pressure. The static magnetic permeability of
a ferrofluid, which depends on the extent of agglomeration, is determined in the
next section.

\section{Impact of agglomeration on the sign of Casimir pressure}

According to the results of Ref.~\cite{33}, the initial susceptibility
of a ferrofluid containing single nanoparticles is given by
\begin{equation}\label{eq5}
   \chi_{\rm ff}^{(1)}(0)=N\frac{M_1^2}{3k_BT},
 \end{equation}
\noindent
 where $N=\Phi/V_1$, the volume of each nanoparticle $V_1=\pi d^3/6$ is expressed
 via its diameter $d$, and $M_1$ is the magnitude of a nanoparticle magnetic
 moment. The later is related to the saturation magnetization per
 unit volume $M_1=M_SV_1$, where $M_S$ may take different values for a bulk
 material and for its parts. Specifically, for single-domain magnetic nanoparticles
 of spherical shape considered here one has
$M_S\approx300~\mbox{emu/cm}^3=3\times 10^5~$A/m
\cite{47}.

Let us now assume that as a result of agglomeration the share $\kappa_k$ of all
nanoparticles is merged into clusters containing $k$ particles each. Taking into
account that the size of clusters is far less than the size of magnetic domain,
it would appear reasonable to put the magnitude of the magnetic moment of a cluster
equal to $M_k=kM_1$. In doing so, the ferrofluid contains $N(1-\kappa_k)$ single
magnetic nanoparticles having the magnetic moments of magnitude $M_1$ and
$N\kappa_k/k$ clusters with magnetic moments of magnitude $M_k$.
Then, the initial susceptibility of a ferrofluid of this type takes the form
\begin{eqnarray}
\chi_{\rm ff}^{(k)}(0)&=&N(1-\kappa_k)\frac{M_1^2}{3k_BT}+
\frac{N\kappa_k}{k}\frac{M_k^2}{3k_BT}
\nonumber \\
&=&[1+(k-1)\kappa_k]\,\chi_{\rm ff}^{(1)}(0),
\label{eq6}
\end{eqnarray}
\noindent
where $\chi_{\rm ff}^{(1)}(0)$ is defined in Eq.~(\ref{eq5}).
As a result, the static magnetic permeability of a ferrofluid allowing for the
effect of agglomeration of nanoparticles is equal to
\begin{equation}\label{eq7}
  \mu_{\rm ff}^{(k)}(0)=1+4\pi\chi_{\rm ff}^{(k)}(0).
\end{equation}

Now we are in a position to compute the Casimir pressure between two parallel plates
separated by a ferrofluid with due account for the effect of agglomeration of
magnetic nanoparticles. We begin with the case of an Au plate described using an
extrapolation of the optical data to low frequencies by means of the plasma model
and a SiO$_2$ plate. Computations are performed by substituting the dielectric
permittivities $\varepsilon^{(1)}$ of Au, $\varepsilon^{(2)}$ of SiO$_2$ and
$\varepsilon_{\rm ff}$ of ferrofluid discussed in Sec.~II to Eqs.~(\ref{eq1})--(\ref{eq4}).

The magnetic permeability of a ferrofluid with account of the effect of agglomeration is
found from Eqs.~(\ref{eq6}) and (\ref{eq7}). Thus, if magnetite nanoparticles have
$d=10~$nm diameter, one obtains $\mu_{\rm ff}^{(1)}(0)=1.24$ if all nanoparticles are
single and  $\mu_{\rm ff}^{(2)}(0)=1.36$ and $\mu_{\rm ff}^{(3)}(0)=1.48$ if half of
all nanoparticles are merged into clusters by two and three particles, respectively.
In a similar way, if the nanoparticle diameter is $d=20~$nm, one obtains from
Eqs.~(\ref{eq6}) and (\ref{eq7}) $\mu_{\rm ff}^{(1)}(0)=2.9$,
$\mu_{\rm ff}^{(2)}(0)=3.85$, and $\mu_{\rm ff}^{(3)}(0)=4.8$ for the cases when
all nanoparticles are single and when one-half of them  are merged into clusters
by two and three, respectively.

Here and below all computations are performed at room temperature $T=300~$K for
a water-based ferrofluid containing $\Phi=0.05$ volume fraction of magnetite nanoparticles.

\begin{figure}[b]
\vspace*{3cm}
\centerline{\hspace*{2.cm}
\includegraphics{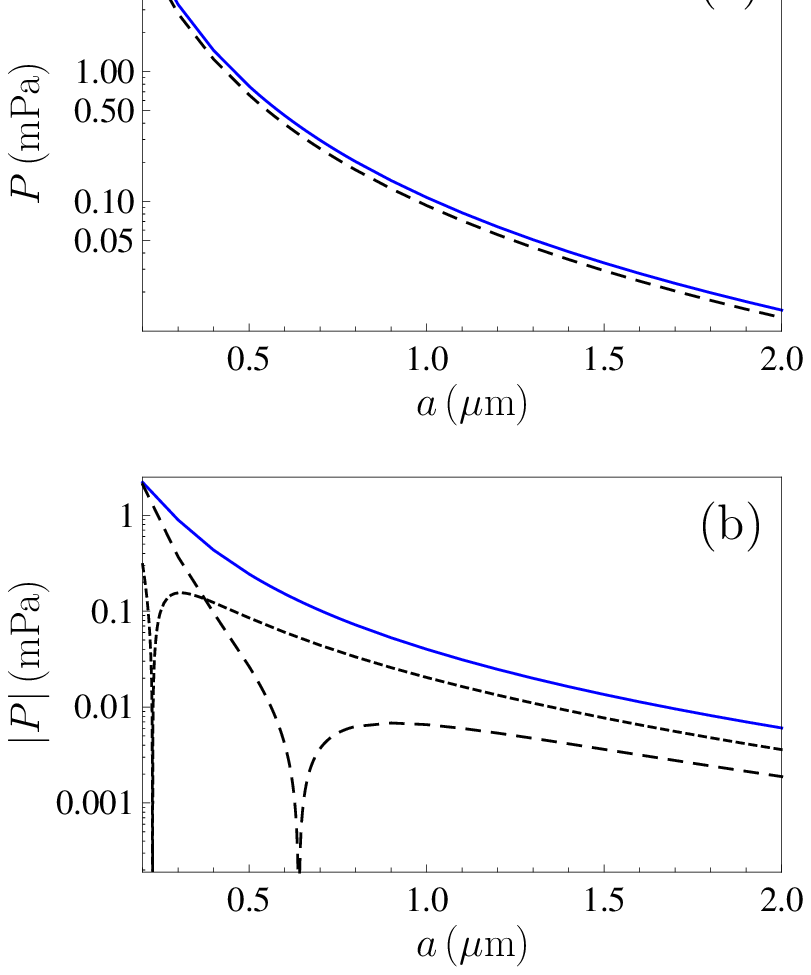}
}
\vspace*{-18.5cm}
\caption{\label{fg1}
(a) The Casimir pressure between Au and SiO${}_2$ plates through
a water-based ferrofluid with 5\% volume fraction of magnetite
nanoparticles and (b) its magnitude are shown as functions of
separation between the plates for nanoparticle diameters (a)
$d = 10~$nm and (b) $d = 20~$nm. The solid, short-dashed and
long-dashed lines are plotted for single nanoparticles of each
type and for the cases when half of them is merged into clusters
by two and three, respectively. Au is described using an
extrapolation of the optical data by means of the plasma model.
}
\end{figure}

In Fig.~\ref{fg1}(a) the computational results for the Casimir pressure are shown
as functions of separation between the plates for nanoparticles of $d=10~$nm diameter
by the solid and dashed lines obtained with no agglomeration and when half of nanoparticles
is merged into clusters by three ($\kappa_3=0.5$), respectively.
As is seen in Fig.~\ref{fg1}(a), for nanoparticles of $d=10~$nm diameter the Casimir
pressure is repulsive and agglomeration leads to only minor quantitative defferences
in the pressure values. Computations show that if one-half of particles were merged
 into clusters by two, the respective line in Fig.~\ref{fg1}(a) would be sandwiched
 between the solid and dashed lines.

In Fig.~\ref{fg1}(b) the computational results for the magnitude of the Casimir pressure
are shown for magnetite nanoparticles of $d=20~$nm diameter. The
solid, short-dashed and long-dashed lines are plotted for the  cases when all nanoparticles
are single, and when half of them is merged
into clusters by two and by three, respectively.
As is seen in Fig.~\ref{fg1}(b), for larger nanoparticles the agglomeration not only
changes the magnitude of the Casimir pressure significantly, but also leads to a
qualitatively different picture  by replacing a repulsion with an attraction when
separation between the plates decreases. Thus, for $\kappa_2=0.5$ the Casimir force
becomes attractive at $a<228~$nm (the short-dashed line) and for
$\kappa_3=0.5$ at $a<640~$nm (the long-dashed line).

A profound effect of agglomeration on the Casimir pressure for sufficiently large
nanoparticles finds simple physical explanation. The point is that the agglomeration
does not influence on the dielectric permittivity of a ferrofluid, but makes an
impact only on $\mu_{\rm ff}(0)$. For small $d$, this impact is also rather small and
increases with increasing $d$ (see above). In the Lifshitz formula (\ref{eq1}),
all terms with $l\geq 1$ lead to an attraction, as well as the
TE contribution to the term with $l=0$ (in the system under consideration the latter
is not equal to zero only in the presence of magnetic properties of
intervening liquid).
As to repulsion, it is produced by the TM contribution to the term with $l=0$.
With increasing $d$, the permeability $\mu_{\rm ff}(0)$ quickly increases, and
the combined effect of the terms with $l\neq 0$ and the TE contribution to the
term with $l=0$ causes a transition from repulsion to attraction at sufficiently
short separations between the plates.

\begin{figure}[b]
\vspace*{-4cm}
\centerline{\hspace*{2.5cm}
\includegraphics{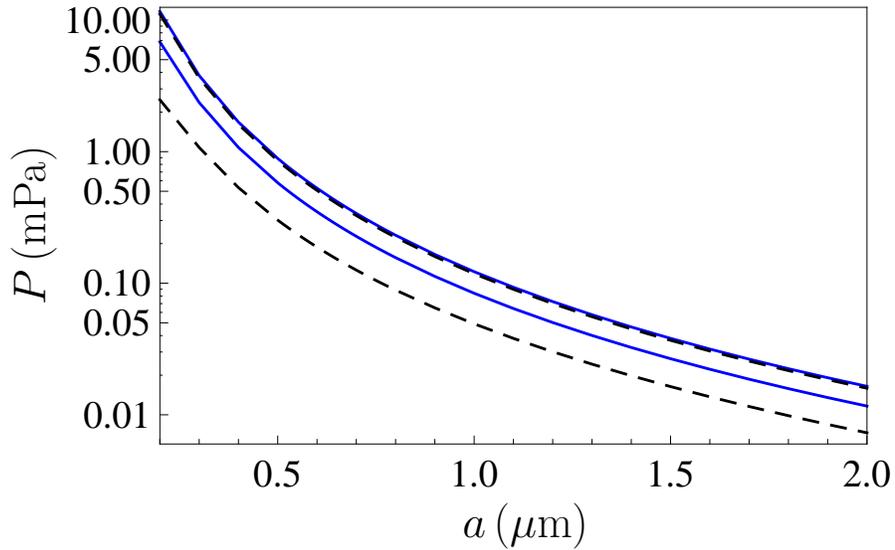}
}
\vspace*{-9.5cm}
\caption{\label{fg2}
The Casimir pressure between Au and SiO${}_2$ plates through
a water-based ferrofluid with 5\% volume fraction of magnetite
nanoparticles is shown as a function of separation between the
plates by the top and bottom pairs of lines for nanoparticle
diameters $d = 10$ and 20~nm, respectively. In each pair, the
solid and dashed lines are plotted for single nanoparticles of
each type and for the case when half of them is merged into
clusters by three, respectively. Au is described using an
extrapolation of the optical data by means of the Drude model.
}
\end{figure}

In computations of Fig.~\ref{fg1}, an extrapolation of the optical data for Au to
low frequencies by means of the plasma model was used. Now we repeat the same
computations but using an extrapolation of the same data by means of the Drude
model. The computational results for the Casimir pressure through a ferrofluid as
functions of separation between the plates are shown in Fig.~\ref{fg2} by the top
and bottom pairs of solid and dashed lines obtained for nanoparticles with
$d=10~$nm and $d=20~$nm diameter, respectively.
In each pair, the solid line is for single nanoparticles and the dashed line is
for the case when half of them is merged
into clusters by  three. By contrast to  Fig.~\ref{fg1}, in Fig.~\ref{fg2} no
qualitative effect caused by the agglomeration of nanoparticles is observed even
for nanoparticles with $d=20~$nm diameter. This is physically explained by the
fact that for typical values of $\mu_{\rm ff}(0)$ calculated above the TE
contribution to the term of the Lifshitz formula with $l=0$
obtained using the Drude model
is much less than that obtained using the plasma model.

\begin{figure}[b]
\vspace*{-4cm}
\centerline{\hspace*{2.5cm}
\includegraphics{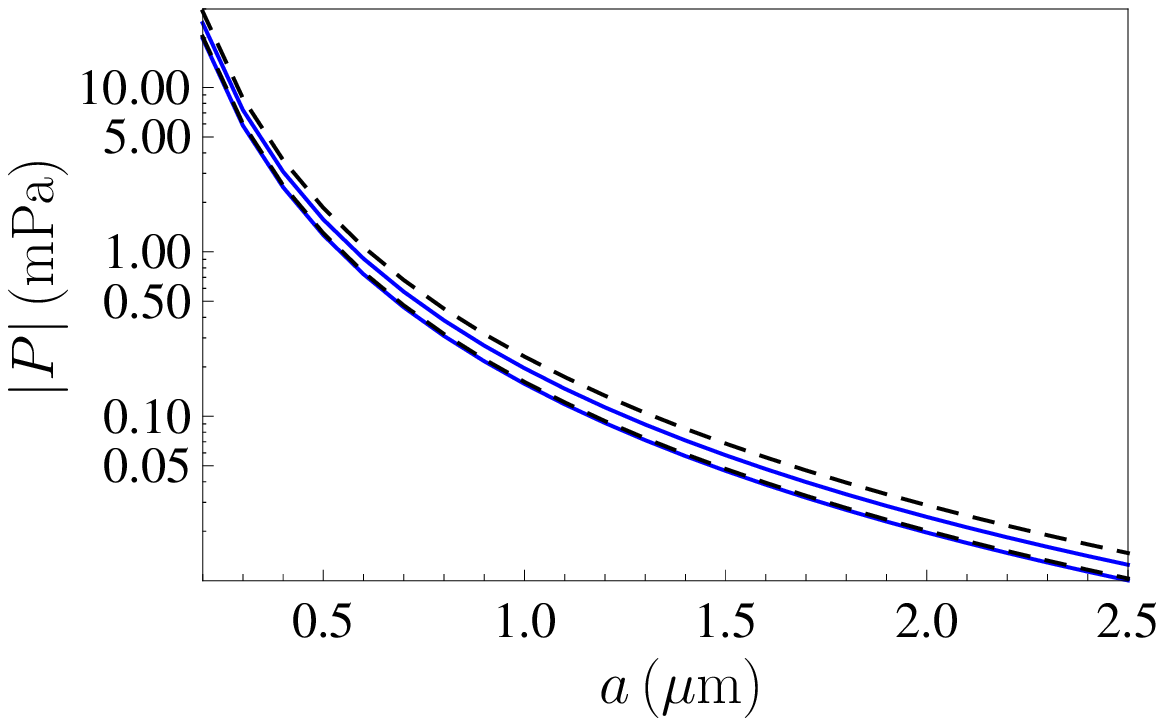}
}
\vspace*{-9.5cm}
\caption{\label{fg3}
The magnitude of the Casimir pressure between two SiO${}_2$
plates through a water-based ferrofluid with 5\% volume fraction
of magnetite nanoparticles is shown as a function of separation
between the plates by the top and bottom pairs of lines for
nanoparticle diameters $d = 20$ and 10~nm, respectively. In each
pair, the solid and dashed lines are plotted for single
nanoparticles of each type and for the case when half of them is
merged into clusters by three, respectively.
}
\end{figure}

For comparison purposes, we also consider the role of agglomeration of nanoparticles
when the ferrofluid is confined between two similar SiO$_2$ plates.
The computational results for the magnitude of the (negative) Casimir pressure are
presented in Fig.~\ref{fg3} by the top
and bottom pairs of solid and dashed lines obtained for nanoparticles with
$d=20~$nm and $d=10~$nm diameter, respectively. As above,
the solid lines are for the case of single nanoparticles and the dashed lines
refer to the case when half of them is merged
into clusters by  three. As is seen in Fig.~\ref{fg3}, for two similar plates
agglomeration of nanoparticles results in only quantitative differences in the
magnitudes of the Casimir pressure. This is explained by the fact that for
similar materials of the plates all contributions to the Lifshitz formula
(\ref{eq1}) add to the effect of attraction.

\section{Dependence on the extent of agglomeration and diameter of magnetic
nanoparticles}

As found in the previous section, the Casimir pressure between metallic and dielectric
plates is subject to change of sign from repulsion to attraction under an impact of
agglomeration of nanoparticles. This effect occurs when the low-frequency dielectric
response of a metal (Au in our case) is described by the experimentally consistent
plasma model (see Sec.~I). Here, we investigate the effect of sign change in relation
to the share of nanoparticles, which are merged into clusters, and nanoparticle
diameter.

We again consider the water-based ferrofluid containing 5\% volume fraction of magnetite
nanoparticles sandwiched between Au and SiO$_2$ plates. For convenience in graphical
displays, we compute the ratio of the Casimir pressures
\begin{equation}
{\cal P}_k=\frac{P(\kappa_k)}{P(\kappa_1)},
\label{eq8}
\end{equation}
\noindent
where ${P(\kappa_k)}$ is the Casimir pressure through a ferrofluid where the share
$\kappa_k$ of all particles is merged into clusters by $k$ particles. Below,
computations of the quantity (\ref{eq8}) are made for $k=2$ and 3.

\begin{figure}[b]
\vspace*{3cm}
\centerline{\hspace*{2.cm}
\includegraphics{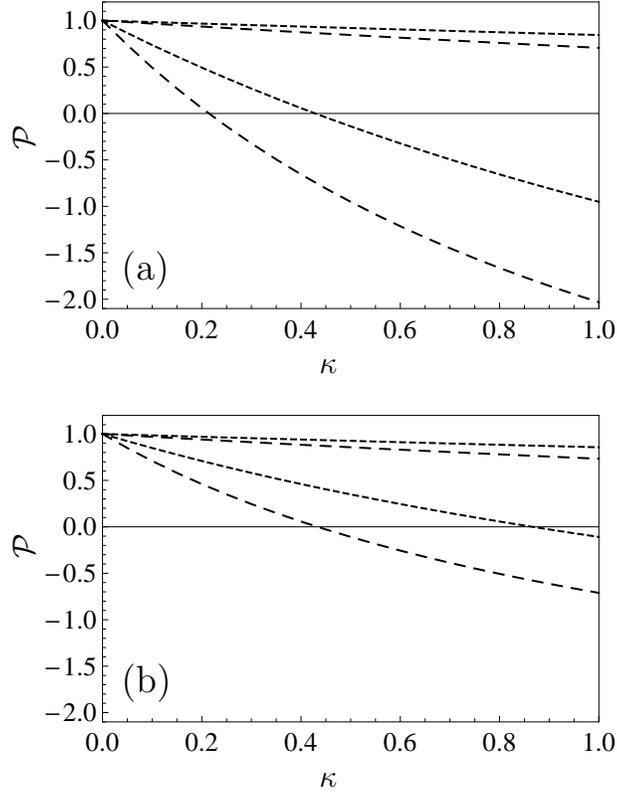}
}
\vspace*{-18.5cm}
\caption{\label{fg4}
The ratio of the Casimir pressure between Au and SiO${}_2$ plates
through a water-based ferrofluid with 5\% volume fraction of
magnetite nanoparticles, of which the share $\kappa_k$ is merged
into clusters by $k$ particles, to the same pressure computed for
single nanoparticles is shown as a function of $\kappa$ at
separation between the plates (a) $a = 200~$nm and (b) $a = 500~$nm,
respectively. In each pair, the short-dashed and long-dashed lines
are plotted for the clusters consisting of $k = 2$ and 3 nanopartices,
respectively. Au is described using an extrapolation of the optical
data by means of the plasma model.
}
\end{figure}

The computational results for ${\cal P}$ as a function of the share of particles
merged into clusters $\kappa=\kappa_2$ or $\kappa_3$ are shown in Fig.~\ref{fg4}(a)
for the plates at $a=200~$nm separation and in Fig.~\ref{fg4}(b)
for  $a=500~$nm. In each of these figures, the top and bottom pairs of lines
are computed for nanoparticle diameters $d=10$ and 20~nm, respectively.
The short- and long-dashed lines label the cases when the share $\kappa$ of all
nanoparticles is merged into clusters by two and three particles, respectively.

As is seen in Figs.~\ref{fg4}(a) and \ref{fg4}(b), for nanoparticles of
$d=10~$nm diameter the effect of sign change does not occur no matter what is
the share of particles merged into clusters. This generalizes the respective
results obtained in Sec.~III. From Figs.~\ref{fg4}(a) and \ref{fg4}(b)
it is also seen that if clusters contain lesser number of particles the
effect of sign change occurs when larger share of all particles is merged
into clusters. Thus, in Fig.~\ref{fg4}(a) the sign change takes place for
$\kappa_2=0.43$ (the short-dashed line) and
$\kappa_3=0.23$ (the long-dashed line). Furthermore, from the comparison of
Fig.~\ref{fg4}(a) and Fig.~\ref{fg4}(b), one can conclude that at larger
separation between the plates the effect of sign change comes for larger
shares of particles merged into respective clusters
(at $a=500~$nm we have $\kappa_2=0.87$ and
$\kappa_3=0.42$ for clusters consisting of two and three particles, respectively).

\begin{figure}[b]
\vspace*{3cm}
\centerline{\hspace*{2.cm}
\includegraphics{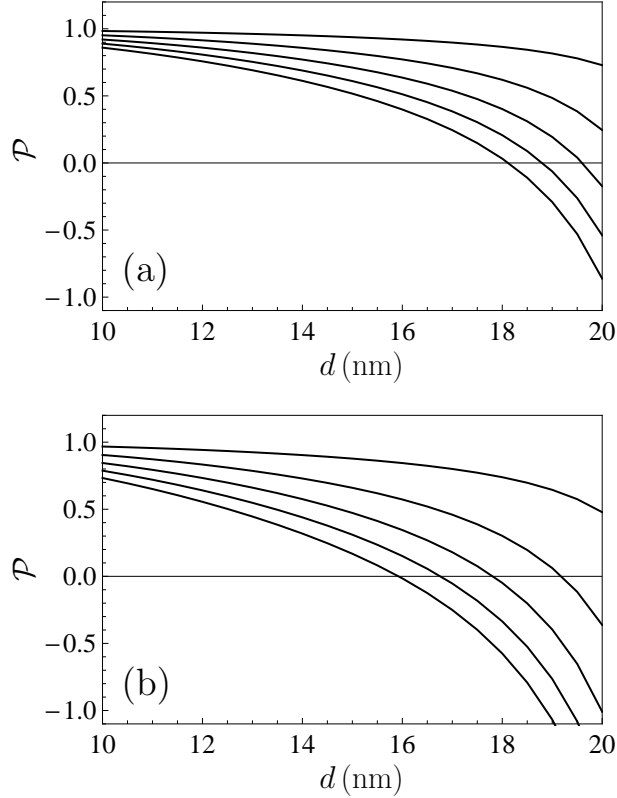}
}
\vspace*{-18.5cm}
\caption{\label{fg5}
The ratio of the Casimir pressure between Au and SiO${}_2$ plates
through a water-based ferrofluid with 5\% volume fraction of
magnetite nanoparticles, of which the share $\kappa_k$ is merged
into clusters by (a) $k = 2$ particles and (b) $k = 3$ particles,
to the same pressure computed for single nanoparticles is shown as
a function of nanoparticle diameter at separation between the plates
$a = 200~$nm. The lines counted from top to bottom are plotted for
the share of merged particles equal to $\kappa_2$ and $\kappa_3 =0.1$,
0.3, 0.5, 0.7, and 0.9, respectively. Au is described using an
extrapolation of the optical data by means of the plasma model.
}
\end{figure}

Finally, we consider how the quantity ${\cal P}$, defined in Eq.~(\ref{eq8}),
depends on the diameter $d$ of single nanoparticles. For this purpose, we
compute ${\cal P}$ as a function of $d$ under different assumptions concerning
the share of merged particles, separation between the plates and the character
of clusters. The computational results for ${\cal P}$ as a function of $d$
at $a=200~$nm are shown in Figs.~\ref{fg5}(a) and \ref{fg5}(b) for the cases
when some share of magnetic nanoparticles is merged into clusters by two and
three, respectively. In each figure, the solid lines counted from top to
bottom are computed for the share of merged particles $\kappa_2$ and
$\kappa_3$ equal to 0.1, 0.3, 0.5, 0.7, and 0.9, respectively.

As is seen in Fig.~\ref{fg5}(a), the effect of sign change occurs only for
nanoparticles of sufficiently large diameter. According to this figure,
the sign of the Casimir pressure changes for
$d>18.1$, 18.7, and 19.6~nm if the shares $\kappa_2=0.9$, 0.7, and 0.5 of all
nanoparticles are merged into clusters by two. Note that we do not consider
nanoparticles with more than 20~nm diameter because otherwise it would be necessary
to increase the minimum separation distance between the plates.
Comparing Figs.~\ref{fg5}(a) and \ref{fg5}(b), one can conclude that if some
share of magnetic nanoparticles is merged into the larger clusters by three
particles each the effect of sign change occurs starting from lesser nanoparticle
diameters. Thus, from Fig.~\ref{fg5}(b) we find that the Casimir pressure
changes its sign from repulsion to attraction for $d>15.9$, 16.7, 17.8, and
19.2~nm if the following respective shares of all nanoparticles are merged
into clusters by three: $\kappa_2=0.9$, 0.7, 0.5, and 0.3.
This opens opportunities to control the sign of the Casimir pressure through
a ferrofluid by choosing an appropriate nanoparticle diameter.
%%%%%%%%%%%%%%%%%%%%%%%%%%%%%%%%%%%%%%%%%%%%%%%%%%%%%%%%%%%%%%%%%%%%%%%%%%%

\section{Conclusions and discussion}

In the foregoing, we have investigated an impact of agglomeration of
magnetic nanoparticles on the Casimir pressure in the configuration
of a ferrofluid sandwiched between two material plates. 
The nanoparticles should be ferromagnetic and their size is restricted 
by the size of one domain for the material under consideration. 
To determine the role of agglomeration,  one needs to know the specific 
magnetic properties of nanoparticles. The most important one is the initial 
magnetic susceptibility of a single nanoparticle which is usually different 
from that determined for bulk material. 
Both cases of
two similar (dielectric) and dissimilar (one dielectric and another
one metallic) plates were considered. Computations of the Casimir
pressure through a ferrofluid have been performed at room temperature
using the Lifshitz theory for Au and SiO${}_2$ plates and the water-based
ferrofluid containing 5\% fraction of magnetite nanoparticles with
different diameters. It was assumed that some share of this nanoparticles
is merged into clusters containing two or three particles. The dielectric
response of Au was described using the measured optical data extrapolated
to low frequencies by means of either the lossless plasma or the lossy
Drude model.

According to our results, for a ferrofluid sandwiched between two
dielectric plates, as well as between dielectric and metallic plates if
the optical data of the latter are extrapolated by means of the Drude
model, an agglomeration of magnetic nanoparticles makes only a
quantitative impact on the Casimir pressure depending on nanoparticle
diameter, but retaining the pressure sign unchanged.

A completely different type of situation occurs for a ferrofluid
sandwiched between one metallic and one dielectric plates when the
low-frequency response of a metal (Au) is described by the plasma
model. In this case, for a sufficiently large nanoparticle diameter,
the agglomeration results in the sign change of the Casimir pressure
from repulsive to attractive. As an example, for magnetite
nanoparticles of 20 nm diameter, half of which is merged into clusters
by three, the pressure becomes attractive at separations between the
plates exceeding 640 nm. It should be taken into account that
numerous experiments of Casimir physics are consistent with the
theoretical predictions using an extrapolation by means of the
plasma model and exclude with certainty the theoretical results
obtained with the help of the Drude model (see Sec. I). Because of
this, the change of the pressure sign as a result of agglomeration of
nanoparticles can be considered as a quantitative effect which merits
detailed consideration.

Based on this conclusion, we have investigated an impact of
agglomeration of nanoparticles on the change of sign of the Casimir
pressure when the share of nanoparticles merged into clusters of
different size and nanoparticle diameter vary continuously. It was
found that the effect of sign change under an impact of agglomeration
becomes more pronounced at shorher separations between the plates, for
larger clusters and arises only for sufficiently large nanoparticle
diameter. To take one example, if 70\% of all nanoparticles are merged
into clusters by three, the Casimir pressure between Au and SiO${}_2$
plates at a distance 200 nm changes its sign if nanoparticle diameter
exceeds 16.8 nm. 
The proposed effects can be observed experimentally in microdevices 
exploiting ferrofluids \cite{14,15,16,17} and in measurements of the 
Casimir force through a liquid layer \cite{liq1,liq2}  when the latter 
possesses magnetic properties.  In doing so, it is simple to generalize 
all the above results for the plates made of any materials if the dielectric 
properties of these materials are available.

To conclude, the obtained results can be used to predict the effect of
agglomeration of magnetic nanoparticles on the Casimir pressure in
microdevices exploiting ferrofluids for their functionality. The
revealed difference regarding the predicted effect of sign change when
using two alternative extrapolations of the optical data of metals to
low frequencies may be of interest for further investigation of the
Casimir puzzle which as yet awaits for its resolution.

\section*{Acknowledgments}

V.~M.~M.~was partially funded by the Russian Foundation for Basic
Research, Grant No. 19-02-00453 A. His work was also partically
supported by the Russian Government Program of Competitive Growth
of Kazan Federal University.

%%%%%%%%%%%%%%%%%%%%
%\end{document}
%%%%%%%%%%%%%%%%%%%%%%__FIGURE_1_%%%%%%%%%%%%%%%%%%%%%%%%%%%%%%%%
%%%%%%%%%%%%%
%%%%%%%%%%%%%%%%%%%%%%__FIGURE_2_%%%%%%%%%%%%%%%%%%%%%%%%%%%%%%%%
%%%%%%%%%%%%%
%%%%%%%%%%%%%%%%%%%%%%__FIGURE_3_%%%%%%%%%%%%%%%%%%%%%%%%%%%%%%%%
%%%%%%%%%%%%%
%%%%%%%%%%%%%%%%%%%%%%__FIGURE_4_%%%%%%%%%%%%%%%%%%%%%%%%%%%%%%%%
%%%%%%%%%%%%%
%%%%%%%%%%%%%%%%%%%%%%__FIGURE_5_%%%%%%%%%%%%%%%%%%%%%%%%%%%%%%%%
%%%%%%%%%%%%%
\end{document}